\documentstyle[prl,aps,multicol,epsf]{revtex}    
\begin{document}    
    
\draft   
 
\twocolumn[ 
\hsize\textwidth\columnwidth\hsize\csname@twocolumnfalse\endcsname 
  
\title{The Current-Temperature Phase Diagram of Layered Superconductors }    
    
\author{ Stephen W.~Pierson\cite{email}}    
\address{Department of Physics, Worcester Polytechnic Institute (WPI), Worcester,    
MA 01609-2280}    
   
\date{Received November 4, 1996}    
\maketitle    
\begin{abstract}    
The behavior of clean layered superconductors in the presence of a finite electric current and in 
zero-magnetic field behavior is addressed. The structure of the current temperature phase 
diagram and the properties of each of the four regions will be explained. We will discuss the 
expected current voltage and resistance characteristics of each region as well as the effects of 
finite size and weak disorder on the phase diagram. In addition, the reason for which a weakly 
non-ohmic region exists above the transition temperature will be explained. 
 
\end{abstract}    
] 
   
\narrowtext    
\section{Introduction}    
\label{sec:intro}  
The understanding of a current's effect on the behavior of vortices in clean layered superconductors 
is important on it owns merit and is essential for the interpretation of measurements which use 
an electric current as a probe. This issue has been addressed by many authors. Glazman and 
Koshelev\cite{glazman90} found that in contrast to quasi-two-dimensional superconducting 
films, there is a non-zero critical current because a minimum current is needed to overcome 
the attraction due to the Josephson coupling between the layers and to drag apart vortex pairs. 
Jensen and Minnhagen\cite{jenmn} have generalized the two-dimensional current voltage 
($I$-$V$) relation to the layered case by considering the effect of the Josephson coupling on 
the intralayer vortex attraction. They find that the voltage goes from being exponentially small 
to finite with the following nonlinear dependence on current: 
\begin{equation} 
V=k(T)I(I-I_c^1(T))^{\alpha(T)} 
\label{IV} 
\end{equation} 
where $I_c^1(T)$ represents the temperature dependent threshold current needed to overcome 
the Josephson attraction and $k(T)$ and $\alpha(T)$ are generally taken to be independent of 
current. The relevance of this equation to such layered materials as the high-temperature 
superconductors (HTSC's) has been established by several experimental 
groups.\cite{balestrino95,miu95a,frelt}  
 
A rigorous self-consistent study of the effects of a current on the critical behavior of vortices 
in a clean layered system was conducted by the author using a renormalization group analysis which 
culminated in the $I$-$T$ phase diagram.\cite{pierson95b} As will be described below, the 
$I$-$T$ space was found to be divided into four regions separated by three characteristic 
currents which included a second order phase transition line. This too has been shown to 
describe well the behavior of the copper oxides.\cite{pierson96b,katona96,pavel96} 
Martynovich and Artemov\cite{artemov96} have also studied this system in zero and finite 
current and also find a second-order phase transition, but only at larger currents. Langevin 
simulations have also been used to study the effect of a driving current on an anisotropic 
Josephson junction array.\cite{dominguez96} Gr{\o}nbech-Jensen,  
Dom\' inguez, and Bishop confirmed the structure of the $I$-$T$ phase 
diagram\cite{pierson95b} finding that the phase transition occurs at a temperature higher than 
the temperature marking the onset of resistance and that the slopes of two of the characteristic 
currents ($I_c^1(T)$ and $I_c(T)$) were linear. While the theoretical and simulational work 
seems sufficient for describing clean, layered systems, there is now evidence that finite size 
effects can produce a linear current dependence in the $V$-$I$ relations at small current 
which give way to a nonlinear behavior at larger 
currents.\cite{katona96,paracchini91,paracchini96,nojima96} This has also been seen in 2D 
systems.\cite{repaci95,repaci96} These and other properties of layered superconductors in the 
presence of a current need to be addressed. 
   
In this paper, we explain in more detail and expand upon the results of 
Ref.~\onlinecite{pierson95b}. In particular the details of how the $I$-$T$ phase diagram was 
arrived at (Section \ref{sec:modelB}), the nature of the actual phase transition (Section 
\ref{sec:modelC}), and the properties of each region (Section \ref{sec:body}) are explained 
more thoroughly. The origin of a weak non-ohmic contribution above the transition 
temperature will be discussed in Sections \ref{sec:bodyC} and \ref{sec:bodyD} as will the 
effects of free vortices created due to finite size effects (Section \ref{sec:finite}). Finally, the 
effects of weak disorder as calculated via the replica technique will be reported (Section 
\ref{sec:disorder}).  
 
\section{Model and derivation of $I$-$T$ phase diagram}    
\label{sec:model} 
 
\subsection{Model and Recursion Relations} 
\label{sec:modelA}

We begin with a description of our model which has been described in detail  
elsewhere\cite{pierson94b,pierson95a} and so will be kept brief here. Vortices in layered 
systems are modeled as a ``gas'' of charges in a stack of weakly coupled layers interacting via 
two-body interactions  
which approximate those of vortices in layered superconductors. Interactions between vortices 
in the  
same layer and between vortices in neighboring layers are included but interactions between 
vortices  
separated by more than one layer are neglected.  The effect of the current ${\bf J}$ is to exert 
a constant  
force on vortices in a direction perpendicular to ${\bf J}$ but in opposite directions for 
oppositely charged  
vortices.  Knowing the interaction potentials and the effect of the current, one can write down 
the partition  
function and perform a real-space renormalization group study on it in the fashion of  
Kosterlitz.\cite{kost} The results are,  
\begin{equation} 
\label{rra} 
dx/d\epsilon=2y^2[1-(1/16)\lambda+J^2], 
\end{equation} 
\begin{equation} 
\label{rrb} 
dy/d\epsilon=2y[x+(1/2)\lambda\ln\lambda]/(1+x) +Jy, 
\end{equation} 
\begin{equation} 
\label{rrc} 
d\lambda/d\epsilon=2\lambda[1-4y^2(1+J^2)/(1+x)], 
\end{equation} 
\begin{equation} 
\label{rrd} 
dJ/d\epsilon=J. 
\end{equation} 
$\lambda$ is the ratio of the interlayer coupling to the intralayer coupling $p^2$, 
$\epsilon=\ln(l/\xi_0)$  
(where $l$ is the relevant length scale),  $x=4/(\beta p^2)-1$, and $\xi_0$ is the zero-
temperature  
correlation length. $y=\exp(-\beta E_c)/l^2$ is the fugacity, where $E_c$ is the ``core energy.'' 
Note that  
the current affects not only the interaction strength parameters $x$ and $\lambda$ but also the 
fugacity  
directly. This is because the current lowers the core energy of vortices and therefore directly 
affects the  
fugacity. 
 
\subsection{The Correlation Length and the $I$-$T$ phase diagram} 
\label{sec:modelB} 
 
The $I$-$T$ phase diagram is derived by studying the correlation length $\xi(T)$ which is 
related to  
the value of $l$ at which one terminates the integration of the recursion relations. Because the 
calculation leading to Eqs.~(\ref{rra})-(\ref{rrd}) assumed small vortex density ($y$), current, 
and interlayer coupling, the integration of the recursion  
relations must be terminated when any of these quantities becomes large. The value of $l$ at 
the cutoff is  
called $l_{max}$ and because only one length (the correlation length) dominates the critical 
behavior, the  
following identification follows: $l_{max}=\xi$. The temperature dependence of $\xi$ is 
found by considering the first integral of the 2D recursion relations: $y^2+2\ln(1+x)-2x=c= 
y_i^2+2\ln(1+x_i)-2x_i$, where the subscript $i$ denotes the initial or ``bare'' value of that 
parameter. $c$ is a constant of integration and it can be shown that $c\propto (T- 
T_{KT})/T_{KT}$\cite{kost} where $T_{KT}$ is the 2D transition temperature. Because we 
have assumed the interlayer coupling to be very weak in our model, it is expected that $c$ 
should be linear in temperature in the immediate vicinity of $T_{KT}$ for our system.   
 
\begin{figure}[htb]
\centerline{\epsfysize=2.4truein \epsffile{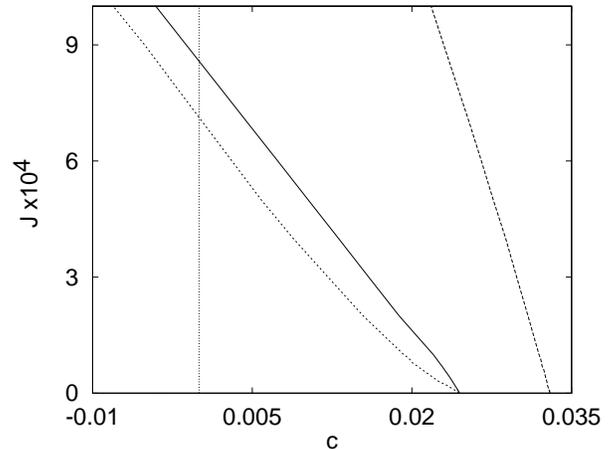}}    
\caption[]{\em{The $I$-$T$ phase space with the three characteristic currents derived by
studying the correlation lengths as explained in the text. The $x$-axis label $c$ is 
prop ortional to $T/T_{KT}-1$.}}
\label{ITPD1}
\end{figure}

Plotting $\ln  (l_{max}/\xi_0)$ versus $c$ for various currents gives an indication of the 
temperature dependence of $\xi(T)$ and the plots are qualitatively similar to that of Fig.~2 in  
Ref.~\onlinecite{pierson95a}. One finds that there is a peak in $\xi(T)$ which occurs at the 
transition temperature. As one increases the current, the peak shifts to the left corresponding to 
a decrease in the transition temperature. This process allows one to determine $T_c(I)$ (or 
equivalently $I_c(T)$) which decreases linearly with current.  $I_c(T)$ is plotted in 
Fig.~{\ref{ITPD1}} as calculated for $x_i=0.5$ and $\lambda_i=10^{-6}$. ($y_i$ and $J_i$ 
were varied to sweep through $I$-$T$ space.) 
 
Two more characteristic currents can be derived by comparing the correlation length for the 
layered,  
finite-current case (referred to as the full correlation length) to the {\bf 2D} finite-current 
correlation length and to the layered {\bf zero-current} correlation length.  In Fig.~1 of 
Ref.~\onlinecite{pierson95b}, the three correlation lengths are plotted. As expected, the 2D 
and full correlation lengths coincide at larger temperatures signifying the 2D behavior of the 
layered system for temperatures larger than $T_c(I)$. The small temperature range over which 
these two correlation lengths separate is taken to be the 3D/2D crossover region and marks  
another characteristic temperature $T_c^2(I)$ (or equivalently $I_c^2(T)$) which is also 
found to vary linearly with current. (See Fig.~{\ref{ITPD1}}.) 
 
The third characteristic current comes from comparing the full correlation length with the 
layered, zero-current correlation length. Where these two correlation lengths begin to separate 
is the temperature at which the current starts to affect the system and identifies $T_c^1(I)$ (or 
$I_c^1(T)$). Like $I_c^2(T)$, it is a crossover current and not a phase transition. By varying 
the current, it is found that $I_c^1(T)$ is linear at larger currents and is roughly parallel to 
$I_c(T)$ but crosses over to join $I_c(T)$ at small currents. In other words, 
$I_c^1(T_c)=I_c(T_c)$ as illustrated in Fig.~\ref{ITPD1}. As we shall discuss below, what 
happens in finite current at two different temperatures, occurs at one temperature in zero 
current. 
 
While $I_c(T)$ can be determined to some precision, the other two characteristic currents 
cannot be since they are crossover currents. One can determine them in different ways which 
yield results which are in qualitative agreement with each other. One can simply take the 
temperature at which the difference between $l_{max}(J,\lambda)$ and, say,  
$l_{max}(J=0,\lambda)$ is a certain value (or at which this difference is a certain percentage 
of  $l_{max}(J,\lambda)$) to be $T_c^1(I)$. While the values are arbitrary, it is found that the 
temperature dependencies of $I_c^1(T)$ and $I_c^2(T)$ do not depend on the chosen values 
nor on whether one chooses the absolute difference or the relative difference. What does 
depend on these factors are the slopes. For this reason, it should be emphasized that 
Fig.~{\ref{ITPD1}} is only a rendition of the phase diagram. The slopes of 
these lines as well as the temperature and current scales will vary not only from the 
less anisotropic materials to the more anisotropic material but also from sample to 
sample.\cite{katona96} The fact that the temperature scale and the current scale cannot be 
determined in our RG analysis contributes to this ambiguity.\cite{pierson95a} 
 
The temperature dependence of $I_c^1(T)$ has also been considered by Jensen and 
Minnhagen\cite{jenmn} who found it to be linear at all currents\cite{detail} unlike the small-
current behavior found in Ref.~\onlinecite{pierson95b}. Using mean-field theory, they found 
that $dI_c^1/dT\propto 1/\gamma$ where $\gamma=\xi_{ab}/\xi_c$ is the anisotropy factor. 
(The relationship between this quantity and our parameter for the strength of the interlayer 
coupling is $\gamma\propto 1/\sqrt{\lambda}.$)  This issue could not be addressed in our 
approach for the following reason. When the anisotropy is large, $T_c^1<<T_c$ and our 
temperature scale, which is linear in $t=T/T_{KT}-1$, is only valid for small $t$. For similar 
reasons, the dependence of $I_c^1(T)$ on $\gamma$ could not be determined. For $I_c(T)$, 
no systematic dependence on $\gamma$ could be identified. 
 
\subsection{The phase transition at $I_c(T)$} 
\label{sec:modelC} 
 
As mentioned above, a second order phase transition of this system occurs at $I_c(T)$. It is 
here that the non-analyticity in the correlation length occurs and above which $y(l)$ goes to 
infinity for large $l$ and below which it goes to zero. In terms of vortices, we believe that at 
this current, free vortex lines can be spontaneously created as opposed to being created by a 
thermal unbinding of vortex loops. This means that the total vorticity of the system need not 
be zero but will fluctuate around zero. This leads to the meaning of  $I_c^1(T)$ which we see 
as a natural extension of the definition of $I_c(T)$.  As we have stated before, at this current 
the system starts to be affected by the current which we have taken to mean that it is now large 
enough for vortex loops to be thermally unbound into vortex lines. In other words, there are 
two mechanisms for creating vortex lines, one is by an unbinding of vortex loops and the other 
is by spontaneous creation. The latter is  possible because the energy of the vortex line 
becomes finite due to screening. In zero current these two mechanisms become possible at the 
same temperature whereas in finite current, they occur at different temperatures. This is why 
$I_c(T)$ and $I_c^1(T)$ start out at the same point in zero current but become separate lines 
at finite current. As mentioned above, Martynovich and Artemov\cite{artemov96} have 
studied this system in zero and finite current. Without a current, they predict a first-order 
phase transition which turns into a second-order phase transition at a finite current.  No 
evidence for this has been presented however and in fact there is evidence for a second order 
phase transition at zero-field.\cite{schilling96}  
 
Gr{\o}nbech-Jensen, Dom\' inguez, and Bishop\cite{dominguez96} have arrived at 
conclusions similar to ours based on Langevin simulations of anisotropic 3D, current-driven 
Josephson Junctions arrays. There they calculate two quantities, the voltage and the helicity 
modulus. They find that the helicity modulus goes to zero at a temperature above which the 
voltage becomes finite. They conclude that the phase transition occurs at a temperature higher 
than that at which vortex loops start to thermally unbind in agreement with our results. They 
also find that $I_c^1(T)$ and $I_c(T)$ are linear in temperature. 
 
To summarize this section, Fig.~\ref{ITPD1}, where $I_c^1(T)$,  $I_c(T)$, and $I_c^2(T)$ 
are plotted, represents the ``raw'' results of our analysis. $I_c^1(T)$ is the value of the current 
at which the system starts to be affected by the current; $I_c(T)$ is the current at which the 
phase transition occurs; and $I_c^2(T)$ is the current at which the 3D/2D crossover occurs. 
These results were obtained for a system composed strictly of vortex pancakes and neglect the 
structure and behavior of the underlying superfluid. Amplitude fluctuations of the superfluid 
become significant near the mean field transition temperature $T_{c0}$ and the superfluid is 
next to non-existent making vortices a minor detail. Minnhagen\cite{minnhagen87} has taken 
this into account and we will see an example of its effect when we consider Eq.~(\ref{RT}). 
(The position of $T_{c0}(I)$ relative to $T_c^2(I)$ could not be determined in our analysis.) 
In the next section we will relate our results to layered superconductors such as the high-
temperature superconductors. In particular we will describe the expected electric transport 
properties that are heavily influenced by vortices in each region for these materials.  
 
\section{$I$-$V$ and $R(T)$ characteristics of the $I$-$T$ Phase Diagram }    
\label{sec:body}  
\begin{figure}[htb]
\narrowtext
\epsfxsize=2.7truein
\hbox{\hskip 0.00truein
\epsffile{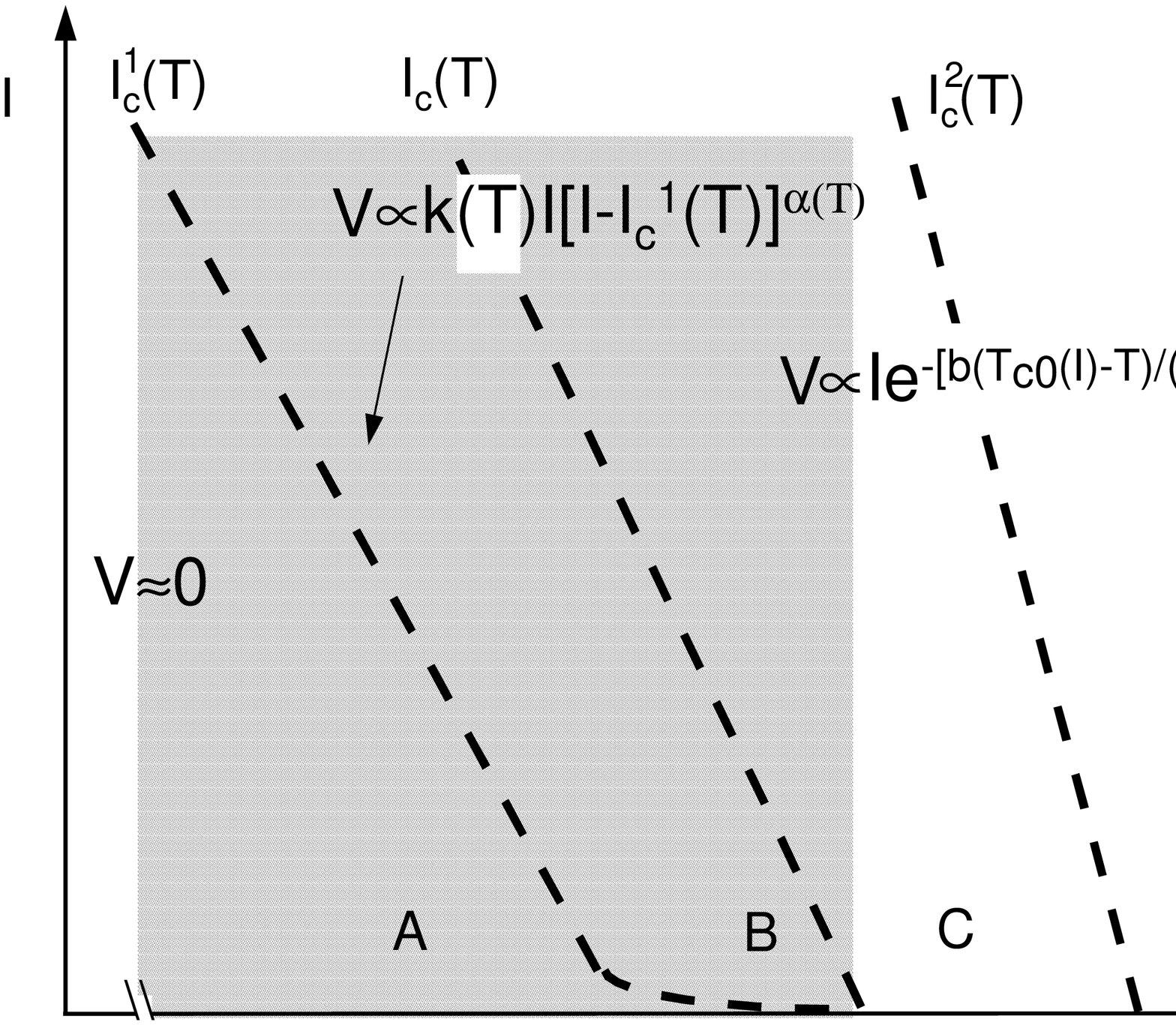}}
\vspace{-0.7truein}
\caption{\em{The expected current-voltage characteristics of the $I$-$T$ phase diagram.
In Region A, $V$ is exponentially small but not zero because, as discussed in
Sec.~\ref{sec:bodyA}, vortex loops can blow out in a direction perpendicular to the
layers. In Region B, Eq.~(\ref{IV}) holds at small currents (shaded area) but will
break down at larger currents. Region D has a weak non-ohmic contribution
(Eq.~(\protect\ref{RT})) which enters through the renormalization of the transition
temperature.}}
\label{ITPD2}
\end{figure}

In this section, the regions of the $I$-$T$ phase diagram will be discussed and the expected 
electrical transport properties (e.g., $I$-$V$ curves and resistance $R(T)$ measurements) of 
each region will be explained. (See Fig.~\ref{ITPD2}.) Flux transformer measurements and 
the detection of $I_c^2(T)$ are also discussed. 
 
\subsection{Region A} 
\label{sec:bodyA}    
This region is defined by where the full correlation length and the layered, zero-current 
correlation length are the same. The coincidence of these two quantitites is interpreted to mean 
that in this region the current does not have a significant effect on the system. In  
the original paper reporting the $I$-$T$ phase diagram,\cite{pierson95b} the current-voltage  
characteristic was taken to be $V(I)=0$. In reality, it should read $V(I)\simeq 0$ since vortex 
loops do have a exponentially small chance of being unbound in the presence of a 
current\cite{glazman90,ffh} because the current also exerts a force on the segments of the 
loop which lie between and parallel to the layers. In a weakly coupled layered system this can 
be neglected and $I_c^1(T)$  signifies a blowout of the loops in a direction predominantly 
parallel to the planes. As one goes to the more isotropic systems, this quantity loses its 
meaning since the loop can blow out in both the parallel and perpendicular directions 
simultaneously. In their simulations, Gr{\o}nbech-Jensen {\it et al.}\cite{dominguez96} do 
find a small number of blow-outs at currents less than $I_c^1(T)$. It should be noted that in 
the two-dimensional limit, this region does not exist because $I_c^1(T;\gamma=\infty)=0$. 
 
\subsection{Region B} 
\label{sec:bodyB}  
In this region, the full correlation length and the layered, zero-current correlation length differ 
from one another signifying that the current has a significant effect on the critical behavior. 
This implies that just above $I_c^1(T)$, vortex loops are being thermally unbound in the 
direction parallel to the layers and the $I$-$V$ relation (Eq.~(\ref{IV})) of Jensen and 
Minnhagen\cite{jenmn} should hold. But this equation should not be expected to hold in all of 
Region B for a number of reasons. First of all, the equation is based upon the formula for rare 
escapes over a barrier.\cite{chandra} This breaks down severely as one increases the current 
and the energy of the most energetic pairs approaches that of the barrier height. Secondly, it is 
based upon vortex loop unbinding in the parallel direction (i.e., 2D unbinding). Close to the 
transition temperature, the system is three-dimensional in its behavior and loop blow outs in 
the perpendicular direction become significant.  These effects may be represented to first 
order by introducing a current dependence into the parameters of  Eq.~(\ref{IV}). In other 
words, to be more strict one should write Eq.~(\ref{IV}) as $V=k(T)I(I-{\cal 
I}(T,I))^{\alpha(T,I)}$ with ${\cal I}(T,I\sim I_c^1(T))\simeq I_c^1(T)$. At larger currents, 
${\cal I}(T,I)$ goes to zero.\cite{gupta93} The behavior of  $\alpha(T,I)$ is less clear.  Our 
renormalization group analysis showed that  $\alpha(T,I)$ should be independent of current 
over a substantial part of region B at fixed temperature but this analysis does not incorporate 
either of the effects mentioned above. What is certain is that $\alpha(T,I)$ goes to zero at 
$I_c(T)$. The approximate region in which Eq.~(\ref{IV}) is expected to hold is represented 
in Fig.~\ref{ITPD2} by the shaded region. 
 
\subsection{Region C } 
\label{sec:bodyC} 
Region C corresponds to temperatures above $T_c(I)$ where the full correlation length differs 
from the 2D finite-current correlation length. Because one is above the phase transition free 
vortex lines can be spontaneously created which one typically considers to mean that the 
region is ohmic. However, as we will show here and in Section \ref{sec:bodyD}, a non-ohmic 
contribution enters through the renormalization of the critical temperature due to the current. 
There are actually non-ohmic contributions from two sources. The first source is of course 
blow-outs of smaller vortex loops which still exist above $I_c(T)$. The other piece will enter 
through the term describing the spontaneously created free vortex lines which is ``largely" 
ohmic. The resistance is proportional to $1/\xi(T)^2$ but a current dependence enters into this 
through the current-dependent transition temperature $T_c(I)$. Because our RG analysis does 
not accurately describe the 3D region, the functional dependence of the resistance in Region C 
cannot be determined. The situation is better in Region D.  
 
A curiosity of this region is that its width increases with current implying that the 3D region 
becomes larger even though the current tends to decouple the layers. We believe that this 
effect is best looked at by saying that the current has a stronger effect on $I_c(T)$ than 
$I_c^2(T)$. It should also be noted that because the full correlation length differs from the 2D 
finite-current correlation length in this region, the behavior should be of a 3D nature. At the 
same time however, the interlayer coupling at large length scales is renormalized to 
zero\cite{pierson95c,pierson96a,friesen95a} making 2D signatures conceivable. 
 
\subsection{Region D } 
\label{sec:bodyD} 
 
In this region, the full correlation length and the 2D finite-current correlation length overlap 
meaning that the behavior of the layered system is 2D-like. One can therefore write down the 
resistance formula for this region based on Minnhagen's 2D formula,\cite{minnhagen87} 
$R(T)=\exp\{-[b(T_{c0}(I)-T)/(T-T_{KT}(I))]^{1/2}\}$, where $ T_{c0}$ is the mean-field 
transition temperature, $T_{KT}$ is the 2D transition temperature and $b$ is a constant. To 
use this for Region D in our the layered system, we must insert the current dependences for the 
two transition temperatures: 
\begin{equation} 
R(T)=\exp\bigl \{-[b(T_{c0}(I)-T)/(T-T_{KT}(I))]^{1/2}\bigr \}, 
\label{RT} 
\end{equation}  
This equation is important because it shows how a weak current dependence can enter into the 
resistance formula making this region weakly non-ohmic. Eq.~(\ref{RT}) has been checked 
for a $YBCO$ data and it is found to work well\cite{paracchini97} for weak currents ($\leq 
10^{-3}\AA$) using a constant $T_{c0}$. It should be stressed that $T_{KT}(I)$ is not the 
same as $T_c(I)$. $T_{KT}(I)$ is the temperature at which it appears that the correlation 
length in the 2D region would diverge, and $T_c(I)$ is the actual temperature at which it 
diverges once the 3D effects take over. As noted above in Section \ref{sec:modelC} one 
should keep in mind that the underlying superfluid is also being destroyed in this vicinity and 
so one must also consider normal state effects which can wash out the effects of vortices.  
 
\subsection{$I_c^2(T)$ and Flux Transformer Geometries} 
\label{sec:bodyE} 
 
Here we will show how our results can be used to explain measurements made in the flux 
transformer geometry\cite{giaever} and explain how one could measure $I_c^2(T)$ which is not easily 
determined from the primarily in-plane Eqs.~(\ref{IV}) and (\ref{RT}). In the flux 
transformer geometry arrangement,\cite{giaever} a current is injected in the top of a sample 
while a secondary voltage $V_s(T)$ is measured on the bottom. Typically, the current is held 
constant while the temperature is varied so that horizontal slices of the $I$-$T$ phase diagram 
can be studied. Such a measurement has been carried out in zero-field by Wan {\it et 
al.}\cite{wan93} where it was found that near the transition temperature a peak appeared in 
$V_s(T)$ for various current values. On the low-temperature side of the peak, $V_s(T)$ rises 
rather abruptly from a value zero in contrast to the high-temperature side where $V_s$ 
descends to zero but but starts to increase before reaching it.  
 
\begin{figure}[htb]
\centerline{\epsfysize=2.4in\epsffile{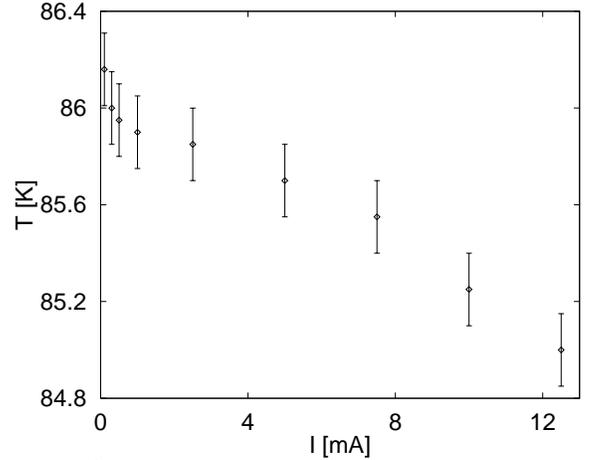}}
\caption{\em{$I_c^1(T)$ determined from $V_s(T)$ as measured in the flux transformer
measurements of Ref.~\protect\onlinecite{wan93}.  As discussed in the text, within 
the error 
bars this data is consistent with our results that $T_c^1(I)$ be linear in current}}
\label{wanic1}
\end{figure}

The behavior of the $V_s(T)$ peak can be explained in terms of the $I$-$T$ phase diagram. 
The onset of the secondary voltage on the low-temperature side is $T_c^1(I)$ where the loops 
starts to blow out due to the current. Because there is finite interlayer coupling in this region, 
the movement of free vortices in the upper layers is translated through to the bottom of the 
sample. The peak gets larger as one goes deeper into Region B because more loops are 
blowing out. However, as one goes into Region C, this effect is offset by the gradual layer 
decoupling and so $V_s(T)$ begins to decrease and in principle will decrease to a value zero 
at $T_c^2(I)$. If one could neglect amplitude fluctuations, $V_s(T)$ would decrease to zero 
at $T_c^2(I)$ but this was not the case in Ref.~\onlinecite{wan93}. We have determined 
$I_c^1(T)$ from the secondary voltage data of Ref.~\onlinecite{wan93} using an onset 
voltage of $0.01\mu$V as our threshold. It is plotted in Fig.~\ref{wanic1} along with its error 
bars which were determined from the temperature resolution used in the measurement of 
$V_s(T)$. Within the error bars it is possible to say that the data is linear in $T$ as found by 
others using in-plane $I$-$V$ measurements\cite{balestrino95,miu95a,frelt,katona96} and 
also predicted in Ref.~\onlinecite{pierson95b} although it is not as convincing as the other 
data.  
 
Even though $T_c^2(I)$ could not be determined from the data of Ref.~\onlinecite{wan93} 
this method remains the best hope for studying $I_c^2(T)$. It may be possible to find a 
sample in which the layers become decoupled before the underlying superfluid breaks down 
and a close inspection of the Wan {\it et al.} data suggests that $T_c^2(I)$ might be 
determined at small currents. It should also be noted that the their data does manifests the 
behavior predicted by our results and discussed in Section \ref{sec:bodyC}: as the current is 
increased, $T_c^2(I)-T_c^1(I)$ increases.

\section{Finite Size Effects}    
\label{sec:finite}    
 
In this Section, we consider finite size effects which make the energy of a single vortex finite 
and therefore make it possible to have free vortices in zero-current at temperatures below the 
critical temperature. The presence of these free vortices destroys the 
Kosterlitz-Thouless-Berezinskii (KTB) transition\cite{ktb,bere71} turning it into a 
crossover. This is because the free vortices screen the vortex pairs more strongly than do 
pairs which in turn makes vortex pair unbinding possible at lower temperatures as 
originally discussed by Kosterlitz and Thouless.\cite{ktb} Nevertheless, it is still 
possible to see KTB behavior as we explain 
below. 

There are two ways in which finite size effects enter. In a chargeless 
superfluid, the bare energy of a 2D, single vortex is $\ln L$ where $L$ is the size of the 
system. In charged superfluids,  the energy of a 2D, single vortex goes as $\ln \lambda_L^2/d 
$ where $\lambda_L$ is the London penetration depth and $d$ is the thickness of the film. 
Usually, however, $L$ and $\lambda_L$ are large in typical samples and thus do not smear 
the transition in a significant way. Beasley, Mooij, and Orlando\cite{beasley79} along with others\cite{turkevich79} pointed this out in the late 1970's. In the presence of a current, 
the principle consequence of a finite $L$ or $\lambda_L$ is to contribute an ohmic term to 
the otherwise purely non-linear (below $T_c$) $I$-$V$ characteristics which can dominate at 
low currents and can wash out the phase transition.  
 
In the HTSC's there is evidence that $\lambda_L$ is small and therefore can significantly 
affect the critical behavior in this system. Indeed, linear $I$-$V$ characteristics at small 
currents have been observed by Paracchini and Romano\cite{paracchini91,paracchini96} on 
YBCO, by Matsuo {\it et al.} on layered conventional superconductors,\cite{nojima96} in the 
author's own analysis\cite{pierson96b,katona96} of $I$-$V$ measurements from various 
groups on YBCO and BSCCO materials, and by Repaci {\it et al.} in ultra-thin YBCuO 
films\cite{repaci95,repaci96} who attributed the small current ohmic behavior to finite size 
effects. Repaci {\it et al.}\cite{repaci95,repaci96} go on to report that low-sensitivity data 
are consistent with a KTB transition. Here, we will address finite size effects on the 
critical behavior and the $I$-$T$ phase diagram. While their effect may ultimately destroy 
the phase transition, our position\cite{ktbtran} is that KTB behavior can still 
be observed and the structure of the $I$-$T$ phase diagram is reflected. (The effect of a 
finite $L$ on 2D Josephson Junction arrays was recently considered theoretically and in 
simulations by Simkin and Kosterlitz\cite{kost97}.) 
 
As mentioned above, finite size effects introduce an ohmic term into the $I$-$V$ 
characteristics which are typically purely non-linear below $T_c(I)$ which can be seen 
through the following simplified derivation. The kinetic equation for the formation of free 
vortices in terms of the density of free vortices $n_F$ is  
\begin{equation} 
dn_F/dt=\Gamma_J(T,J)+\Gamma_{FS}(T)-k_1n_F^2, 
\label{kinetic} 
\end{equation} 
where $\Gamma_J(T,J)$ is the rate at which free vortices are being created by pair unbinding 
due to thermal activation over the barrier made possible by the current, $\Gamma_{FS}(T)$ is 
the rate at which free vortices are created due to finite size effects, and $k_1$ is a constant. It 
is a standard derivation\cite{jenmn} to show that 
\begin{equation} 
\Gamma_J(T,J)\propto (I-I_c^1(T))^{2\alpha_B(T)}, 
\label{rate1} 
\end{equation} 
which corresponds to the rate at which bound vortex pairs are thermally activated over a 
barrier and $\alpha_B(T)$ corresponds to a bare exponent that must be renormalized. The rate 
at which free vortices are spontaneously created is proportional to a Boltzman factor:  
\begin{equation} 
\Gamma_{FS}(T)\propto \exp [-E_{FV}/k_BT], 
\label{rate2} 
\end{equation} 
where $E_{FV}$ is the energy of a free vortex. As pointed out above, Eq.~\ref{rate1} is valid 
for rare escapes over a barrier and also for when the vortex loop blowouts are in a direction 
parallel to the planes.  In that limit, one can take $\Gamma_{FS}(T)$ to be independent of 
current. It is Eq.~\ref{kinetic} that one uses to derive 
Eq.~\ref{IV} in the limit of $E_{FV}=\infty$. 
 
\begin{figure}[htb]
\narrowtext
\epsfxsize=2.8truein
\hbox{\hskip 0.0truein
\epsffile{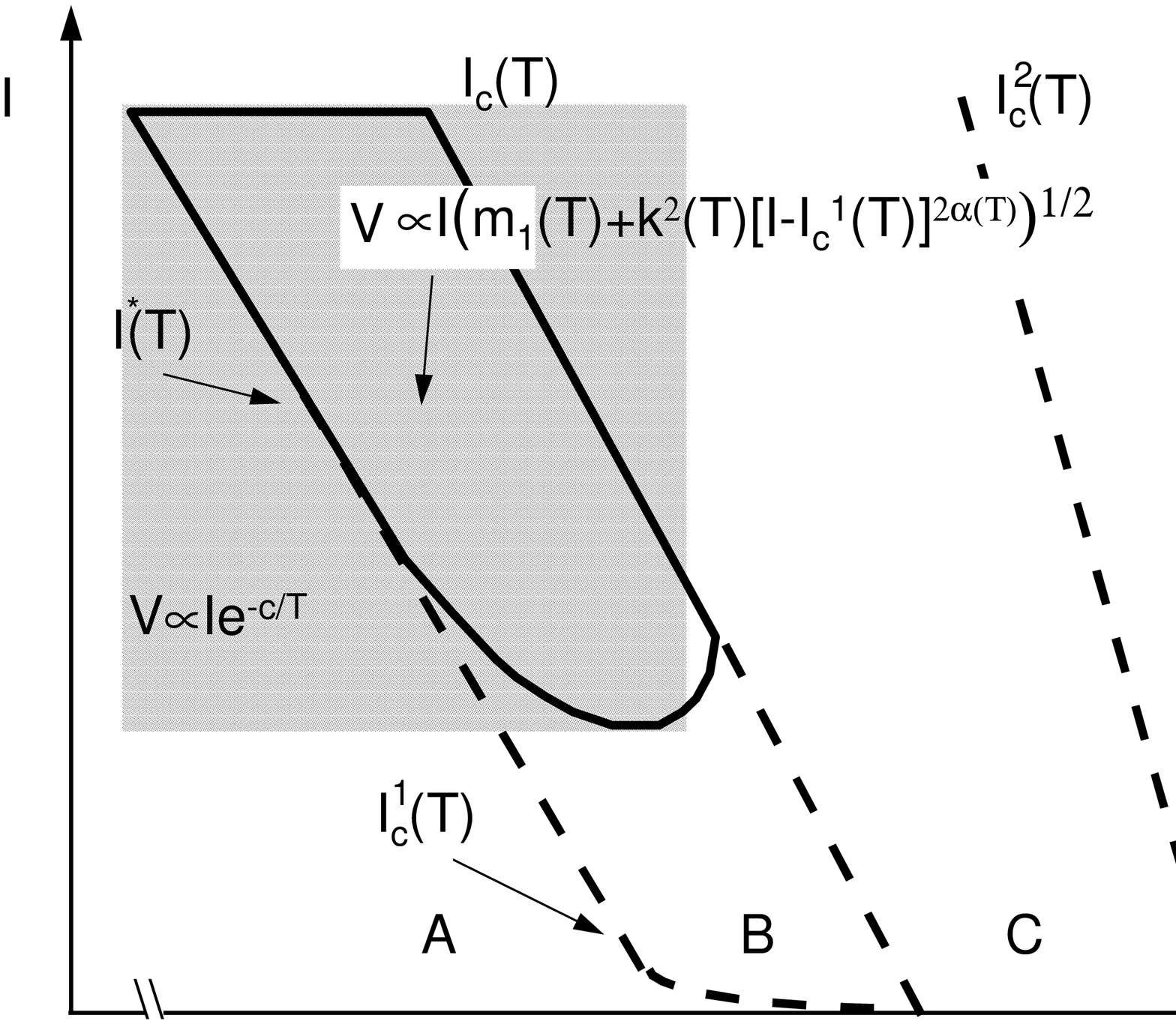}}
\vspace{-0.7truein}
\protect\caption{\em{The approximate modifications to the $I$-$T$ phase diagram due 
to finite-size effects. Note that the original phase lines are shown here as 
dashed lines. Within the solid
lines is where one would expect the effect of vortex-pair unbinding. Outside of this 
region
ohmic behavior is expected although it could be exponentially small at lower 
temperatures and
currents. Slightly non-ohmic behavior may persist to the right on the non-ohmic 
region as a result of transition temperature renormalization.}}
\label{fig3}
\end{figure}

Proceeding with Eq.~\ref{kinetic}, one assumes a steady state solution and finds  
\begin{equation} 
n_F\propto\sqrt{ \Gamma_J(T,J)+\Gamma_{FS}(T)}. 
\label{nffs} 
\end{equation} 
This is readily put in the form of a current-voltage relation since $R\propto n_F$ and $V=IR$: 
\begin{equation} 
V=I\sqrt{m_1(T)+k(T)^2[I-I_c^1(T)]^{2\alpha(T)}}. 
\label{ivfs} 
\end{equation} 
At low currents, the first term will dominate and the $I$-$V$ curves will be linear but at larger 
currents it is possible for the second term to dominate and for the $I$-$V$ curves to become 
non-linear. Therefore, by taking into account the first term, it is still possible to determine 
$I_c^1(T)$ and $\alpha(T)$. The current $I^*(T)$ at which the second term dominates 
introduces a fourth characteristic current into the $I$-$T$ phase diagram which becomes quite 
different in the presence of finite size effects as we discuss below.  If the temperature 
dependence of $m_1(T)$ and $k(T)$ were known, it would be possible to determine the 
relationship between $I^*(T)$ and  $\alpha(T)$. As it is we have, 
\begin{equation} 
I^*(T)=I_c^1(T)+[m_1(T)/k(T)^2]^{1/[2\alpha(T)]} 
\label{istar} 
\end{equation} 
which tells us that $I^*(T)$ lies above $I_c^1(T)$ and that their difference increases with 
temperature. (In this last statement we have made use of some empirical understanding of 
$m_1(T_/k(T)$.) Based on this we can modify the $I$-$T$ phase diagram to include finite-
size effects as we illustrate in Fig.~\ref{fig3}. Within the solid lines is where one would 
expect the non-linear behavior due to vortex pair-unbinding. The lower solid line is $I^*(T)$ 
and the upper is $I_c(T)$. We have not included the effect of the renormalization of $I_c(T)$ 
due to the free vortices here. To a first approximation, their effect will be to shift this quantity 
to lower temperatures and perhaps to smear it. The renormalization of the quantities in 
Eq.~\ref{ivfs} should also be accounted for but here we do not expect any drastic changes. 
Fig.~\ref{fig3} is similar to Fig.~8 of Ref.~\onlinecite{paracchini96} but the differences are 
important. Their $I_0(T)$ is defined through a phenomenological equation different than 
Eq.~(\ref{IV}) and is therefore distinct from $I_c^1(T)$ and $I^*(T)$. Furthermore, the 
ohmic region and non-ohmic regions have different shapes in the two figures.

\section{Effects of Disorder}    
\label{sec:disorder} 
 
In this section, the effect of a quenched, random one-body potential $V_d(R)$ on the system 
will be considered using the replica technique.\cite{pwa75} The primary effect of this type of 
disorder is to pin a vortex, but various types of disorder on 2D systems have been considered 
in the past. Rubinstein {\it et al.}\cite{nelson83} and Korshunov\cite{korshunov93} have 
studied the effects of a random configuration of quenched vortex pairs. Jos\' e\cite{jose81} 
has investigated the 2D ferromagnetic planar model where the coupling is taken to be a 
random variable while Fischer\cite{fischer94} has studied a system where the disorder is 
introduced via the fugacity. The effect of a quenched, random one-body potential on a one-
component vortex gas (i.e., no antivortices)  has been considered by Menon and 
Dasgupta\cite{dasgupta94} and we will follow their notation here. We will present the 
calculation for the layered case which is also applicable to the 2D case. 
 
The essence of the replica technique is to average the disorder over $n$ replicas of the system 
which is done via the identity $[\ln Z]=\lim_{n\rightarrow 0}\int dV_d P(V_d) (Z^n-1)/n$ 
where $Z$ is the partition function for the layered vortex gas and [...] represents an overage 
over the probability distribution of the disorder. $P(V_d)$ is a Gaussian distribution with zero 
mean value and short ranged spatial correlations: $[V_d({\bf r})V_d({\bf r}')] \simeq 
\Delta\delta(|{\bf r}-{\bf r}'|)$ where $\Delta$ represents the strength of the disorder. The 
effect of this averaging is to introduce an interaction between vortices in different replicas 
which includes the standard logarithmic interaction in addition to a short ranged disorder 
induced interaction which is proportional to $-\beta\Delta\delta(|{\bf r}-{\bf r}'|)$ as described 
in more detail in Ref.~\onlinecite{dasgupta94} After this averaging, the model is similar to 
that of Fischer\cite{fischer94} except the additional disorder induced interaction is long 
ranged there. 
 
We have done a renormalization group analysis on the new partition function which consists 
of two steps: a course graining and a rescaling. While the integrations for the course graining 
step cannot be done exactly, one can show that the resulting terms have no logarithmic or 
linear dependencies and so will not renormalize the interaction strength parameters $p^2$ and 
$\lambda$. To do this integral, one approximates the delta function by the short-ranged 
expression,\cite{dasgupta94} $\exp[-r^2/\xi_0^2]$. The rescaling steps alone result in the 
following recursion relation for the disorder strength: $d\Delta/d\epsilon=-\Delta$ which is 
immediately seen by considering $-\beta\Delta\delta(R)$ with the substitution 
$R'=R/(1+d\tau/\tau)$. Since it is unlikely that coarse-graining effects will contribute to 
making the disorder more important, this parameter is irrelevant. 
 
Because the disorder does not affect the recursion relations of the system parameters $p^2$,  
$\lambda$ and $y$ and the disorder parameter $\Delta$ is irrelevant, our results show that a 
weak, random, one-body potential has no effect on the critical behavior of the system in 
agreement with the results of Ref.~\onlinecite{fischer94}. This is expected since even with 
one vortex in a pair pinned, the pair can still unbind because of the ``freedom'' of the other 
vortex. 
 
\section{Discussion and Summary}    
\label{sec:summ} 
 
Numerous aspects of the current-temperature phase diagram have been considered here 
including its derivation, the current-voltage characteristics of each region, properties of the 
phase lines, and the use of the flux transformer geometry to study $I_c^2(T)$. We have also 
considered finite size effects and derived a current-voltage equation for this case. Finally, the 
effect of a weak, random, one body potential was found to be weak. An interesting 
consequence of the current dependence of the critical temperature is to make the system 
slightly non-ohmic above $T_c$. There remain many interesting aspects of the $I$-$T$ phase 
diagram to consider including the dependencies of the characteristic currents $I_c^1(T)$, 
$I_c(T)$, and $I_c^2(T)$ on the anisotropy factor $\gamma$. 
 
\acknowledgements    
The author gratefully acknowledges D. Dom\' inguez, P.~Minnhagen, L.~Miu, C.~Paracchini, 
L.~Romano, and S.~Shenoy, for useful conversations. I also thank S.~Hebboul {\it et al.} for 
providing their $V_s(T)$ data, M.~Friesen for his contribution in deriving Eq.~(\ref{rrb}),  
and T.~M.~Katona for assistance with Fig.~\ref{wanic1}. Early stages of this work were 
carried out at the Naval Research Laboratory and is supported by the Office of Naval 
Research.


\begin{references}
\bibitem{detail} A similar temperature dependence for $I_c^1(T)$ was also found in 
Ref.~\protect\onlinecite{glazman90} if one neglects the temperature dependence of 
$\lambda$.
\bibitem{ktbtran} On the question of whether there is actually a KTB transition in layered 
superconductors depends on how one defines it. If one defines it as an unbinding of vortex 
pairs, then it is unlikely that there is a KTB transition since the transition is known to be 
3D and that loops will dominate the critical behavior in which case the transition will be a 
vortex loop ``blow-out." (One could also characterize the KTB transition by the discontinuity 
in the superfluid density.) In this paper, we say that there is KTB behavior if the 
electrical transport characteristics reflect those generalized from the 2D case: 
Eqs.~(\ref{IV}) and (\ref{RT}).
\end{references}
\end{document}